\documentclass[letterpaper, 10 pt, conference, onecolumn]{ieeeconf}  % Comment this line out if you need a4paper

\IEEEoverridecommandlockouts                              % This command is only needed if 
\usepackage{graphicx} % for pdf, bitmapped graphics files
\usepackage{epsfig}   % for postscript graphics files
\usepackage{mathptmx} % assumes new font selection scheme installed
\usepackage{times}    % assumes new font selection scheme installed
\usepackage{amsmath}  % assumes amsmath package installed
\usepackage{amssymb}  % assumes amsmath package installed
\usepackage{units}    
\usepackage{booktabs} % For table formatting.
\usepackage{cite}     % For bibtex references.
\usepackage{newtxtext, newtxmath}   % To get the usual \mathcal look 
\usepackage{hyperref} % Provides LaTeX the ability to create hyperlinks within and outside of the PDF file. 
\usepackage{float}    % Provides functions to format float placement. 
\usepackage{subfig}   % For breaking figures into subfigures. 

\title{\LARGE \bf
Finite Time Robust Control of the Sit-to-Stand Movement \\
for Powered Lower Limb Orthoses}

\author{Octavio Narvaez-Aroche$^{1}$, Andrew Packard$^{1}$ and Murat Arcak$^{2}$% <-this % stops a space
\thanks{$^{1}$Berkeley Center for Control and Identification, Department of Mechanical Engineering, University of California, Berkeley. {\tt\small ocnaar@berkeley.edu, apackard@berkeley.edu}}%
\thanks{$^{2}$Department of Electrical Engineering and Computer Sciences, University of California, Berkeley, CA, 94720. {\tt\small arcak@berkeley.edu}}%
}

\begin{document}

\maketitle
\thispagestyle{plain}
\pagestyle{plain}

%%%%%%%%%%%%%%%%%%%%%%%%%%%%%%%%%%%%%%%%%%%%%%%%%%%%%%%%%%%%%%%%%%%%%%%%%%%%%%%%
\begin{abstract}

\textit{This study presents a technique to safely control the Sit-to-Stand movement of powered lower limb orthoses in the presence of parameter uncertainty. The weight matrices used to calculate the finite time horizon linear-quadratic regulator (LQR) gain in the feedback loop are chosen from a pool of candidates as to minimize a robust performance metric involving induced gains that measure the deviation of variables of interest in a linear time-varying (LTV) system, at specific times within a finite horizon, caused by a perturbation signal modeling the variation of the parameters. Two relevant Sit-to-Stand movements are simulated for drawing comparisons with the results documented in a previous work. }

\end{abstract}

%%%%%%%%%%%%%%%%%%%%%%%%%%%%%%%%%%%%%%%%%%%%%%%%%%%%%%%%%%%%%%%%%%%%%%%%%%%%%%%%
\section{INTRODUCTION}

In reference \cite{Narvaez-Aroche2017} we proposed a generalizable strategy for planning the Sit-to-Stand (STS) movement of a powered lower limb orthosis for restoring the gait of people with complete paralysis of the lower part of the body and used a controller based on feedback linearization to track two relevant STS maneuvers. Although the simulations exhibited acceptable state tracking errors in the presence of parameter uncertainties, we observed a high sensitivity in the variability of the required control inputs. 

Since the control inputs related to the upper body loads at the shoulders joint are expected to be executed by the user for successfully completing the movements and there is no feedback control/computer authority over them, a key concern was to find a controller that would reduce their deviation from reference trajectories without compromising the desired kinematics for the Center of Mass (CoM), especially in the final standing configuration, in order to decrease the likelihood of observing sit-back or step failures \cite{Eby2006}. To have a robust performance metric for this objective and evaluate different finite time horizon LQR controllers, we opted to compute induced gains for an extended LTV system with a perturbation signal to model the parameter variation as its input, and the deviation from our variables of interest as output. The deduction of such LTV system follows from the nonlinear dynamics model, the reference trajectories obtained from our motion planning strategy and the kinematic equations of the CoM in the following sections. We conclude with the simulation of the same STS movements presented in \cite{Narvaez-Aroche2017} under the control of finite time LQR controllers found through weight selection via our robust performance metric.      
 
\section{DYNAMICS}

\begin{figure}
	\begin{centering}
		\includegraphics[width=4.5cm]{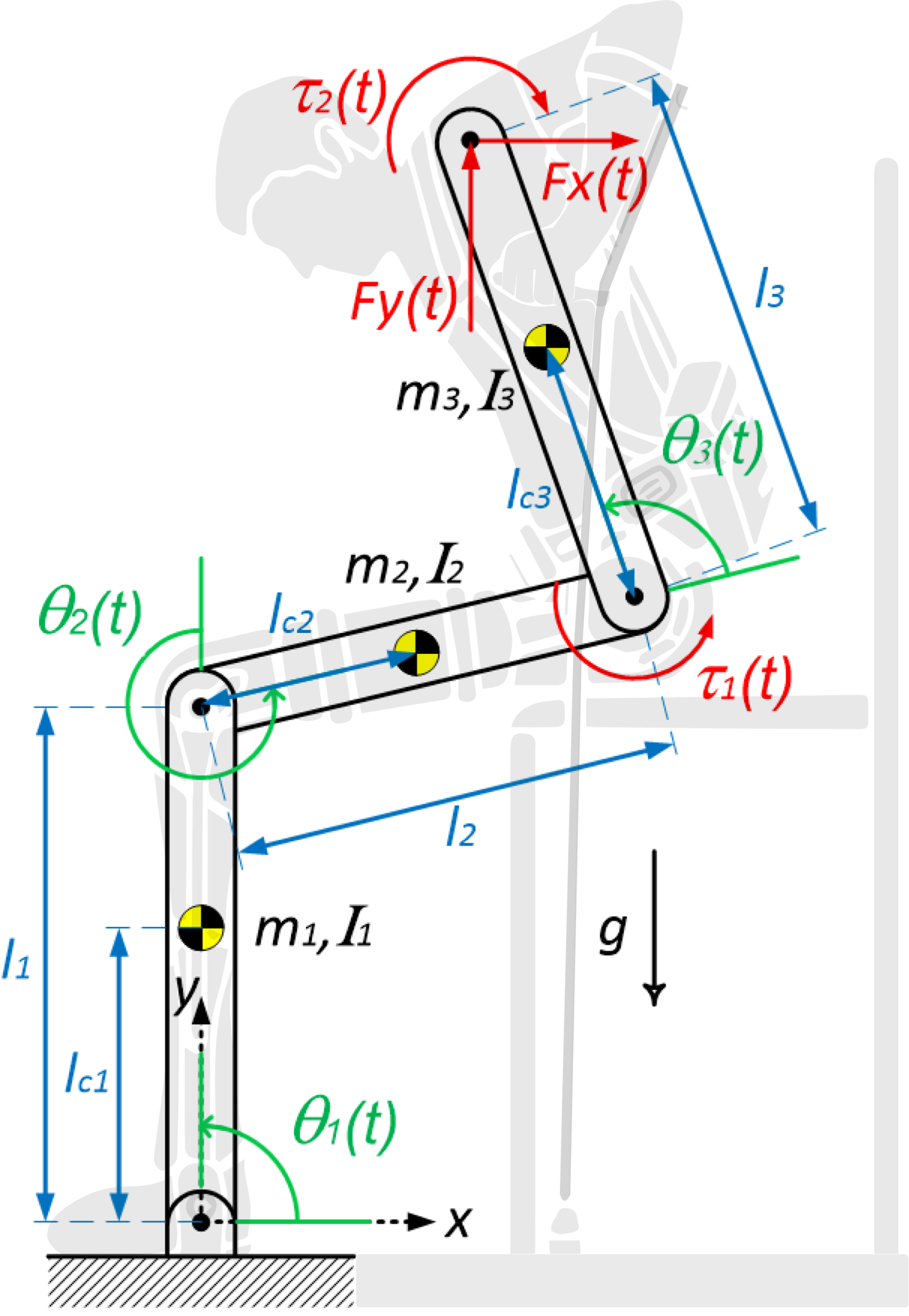}
		\par\end{centering}
	\caption{Three-link planar robot for modeling a powered lower limb orthosis
		and the interaction with its user during a Sit-to-Stand (STS) movement.\label{fig:Robot} }
\end{figure}

Assuming sagittal symmetry, no movement of the head relative to the torso, and that feet are fixed to the ground, we model the user, crutches and powered orthosis as a three-link planar robot with revolute joints coaxial to the ankles, knees and hips, as shown in Fig. \ref{fig:Robot}. $\theta_{1}$ is the angular position of link 1 (shanks) measured from the $x$ axis, $\theta_{2}$ is the angular position of link 2 (thighs) relative to link 1, and $\theta_{3}$ is the angular position of link 3 (torso) relative to link 2. The system parameters are the masses of the links $m_1$, $m_2$, and $m_3$; the moments of inertia about their respective Centers of Mass (CoMs) $I_{1}$, $I_{2}$, and $I_{3}$; their lengths $l_1$, $l_2$, and $l_3$; and the distances of their CoMs from the joints $l_{c1}$, $l_{c2}$, and $l_{c3}$. The actuators of the orthosis exert torque $\tau_{1}$ about the hips; while torque $\tau_{2}$, horizontal force $F_{x}$ and vertical force $F_{y}$ capture the inertial and gravitational forces of the arms and loads applied on the shoulders of the user by its interaction with the ground through crutches. There is no torque applied at the knees in compliance with the architecture of the most affordable medical devices of this kind available in the market \cite{ExoskeletonReportLLC2017,Strickland2016}.
For notational convenience, denote $ c_{i} :=\cos\theta_{i}\left(t\right) $, $ c_{ij} :=\cos\left(\theta_{i}\left(t\right)+\theta_{j}\left(t\right)\right) $, $ c_{ijk} :=\cos\left(\theta_{i}\left(t\right)+\theta_{j}\left(t\right)+\theta_{k}\left(t\right)\right) $,
%{\small{
%		\begin{align*}
%		c_{i} & :=\cos\theta_{i}\left(t\right)\\
%		c_{ij} & :=\cos\left(\theta_{i}\left(t\right)+\theta_{j}\left(t\right)\right)\\
%		c_{ijk} & :=\cos\left(\theta_{i}\left(t\right)+\theta_{j}\left(t\right)+\theta_{k}\left(t\right)\right)
%		\end{align*}}}
%{\small{
%\[
%	c_{i} :=\cos\theta_{i}\left(t\right),
%	c_{ij} :=\cos\left(\theta_{i}\left(t\right)+\theta_{j}\left(t\right)\right),
%	c_{ijk} :=\cos\left(\theta_{i}\left(t\right)+\theta_{j}\left(t\right)+\theta_{k}\left(t\right)\right)
%\]}}P
and similarly for $ \sin\left(\cdot\right) $.
In terms of the joint angles $\theta$, input $u$, parameters $p$: {\small{
\begin{align*}
\theta &=\left[\theta_{1};\;\; \theta_{2};\;\;\theta_{3}\right],\quad \quad \quad u=\left[\tau_{1};\;\; \tau_{2};\;\; F_{x};\;\; F_{y}\right],\\
p&=\left[m_{1};\;\;m_{2};\;\;m_{3};\;\;I_{1};\;\;I_{2};\;\;I_{3};\;\;l_{1};\;\;l_{2};\;\;l_{3};\;\;l_{c1};\;\;l_{c2};\;\;l_{c3}\right],
\end{align*}}}
and
\[
\begin{array}{lll}
k_{0}:=\left(m_{1}+m_{2}+m_{3}\right)^{-1}, &  & k_{1}:=l_{c1}m_{1}+l_{1}m_{2}+l_{1}m_{3},\\
k_{2}:=l_{c2}m_{2}+l_{2}m_{3}, &  & k_{3}:=l_{c3}m_{3},
\end{array}
\]
the Euler-Lagrange equations of the three-link planar robot in Fig. \ref{fig:Robot} can be written, with the aid of the symbolic multibody dynamics package PyDy \cite{Gede2013}, as
{\small \begin{equation}
M\left(\theta\left(t\right),p\right)\ddot{\theta}\left(t\right)+F\left(\theta\left(t\right),\dot{\theta}\left(t\right),p\right)=A_{\tau}\left(\theta\left(t\right),p\right)u\left(t\right). \label{eq:EulerLagrange}
\end{equation}}$M\left(\theta,p\right)\in\mathbb{R}^{3\times3}$, $M\left(\theta,p\right)\succ0$ is the symmetric mass matrix of the system with entries
{\small{\begin{align*}
	M_{11} & =  I_{1}+I_{2}+I_{3}+l_{c1}^{2}m_{1}+m_{2}\left(l_{1}^{2}+2l_{1}l_{c2}c_{2}+l_{c2}^{2}\right)\\
	       & + m_{3}\left(l_{1}^{2}+2l_{1}l_{2}c_{2}+2l_{1}l_{c3}c_{23}+l_{2}^{2}+2l_{2}l_{c3}c_{3}+l_{c3}^{2}\right)\\
	M_{12} & = I_{2}+I_{3}+l_{c2}m_{2}\left(l_{1}c_{2}+l_{c2}\right)\\
	       & + m_{3}\left(l_{1}l_{2}c_{2}+l_{1}l_{c3}c_{23}+l_{2}^{2}+2l_{2}l_{c3}c_{3}+l_{c3}^{2}\right)\\
	M_{13} & = I_{3}+l_{c3}m_{3}\left(l_{1}c_{23}+l_{2}c_{3}+l_{c3}\right)\\
	M_{22} & = I_{2}+I_{3}+l_{c2}^{2}m_{2}+m_{3}\left(l_{2}^{2}+2l_{2}l_{c3}c_{3}+l_{c3}^{2}\right)\\
	M_{23} & = I_{3}+l_{c3}m_{3}\left(l_{2}c_{3}+l_{c3}\right)\\
	M_{33} & = I_{3}+l_{c3}^{2}m_{3}.
\end{align*}}}$F\left(\theta,\dot{\theta},p\right)\in\mathbb{R}^{3}$ is the vector of energy contributions due to the acceleration of gravity $g=9.81\left[\nicefrac{m}{s^{2}}\right]$ and Coriolis forces
{\footnotesize{\[
	F\left(\theta,\dot{\theta},p\right)=\Omega\left(\theta,p\right)\left[\begin{array}{c}
		\dot{\theta}_{1}^{2}\\
		\left(\dot{\theta}_{1}+\dot{\theta}_{2}\right)^{2}\\
		\left(\dot{\theta}_{1}+\dot{\theta}_{2}+\dot{\theta}_{3}\right)^{2}
	\end{array}\right]+g\left[\begin{array}{c}
		k_{1}c_{1}+k_{2}c_{12}+k_{3}c_{123}\\
		k_{2}c_{12}+k_{3}c_{123}\\
		k_{3}c_{123}
	\end{array}\right],
\]}} with 
{\footnotesize \[
\Omega\left(\theta,p\right)=\left[\begin{array}{ccc}
l_{1}\left(k_{2}s_{2}+k_{3}s_{23}\right) & -k_{2}l_{1}s_{2}+k_{3}l_{2}s_{3} & -k_{3}l_{1}s_{23}-k_{3}l_{2}s_{3}\\
l_{1}\left(k_{2}s_{2}+k_{3}s_{23}\right) & k_{3}l_{2}s_{3} & -k_{3}l_{2}s_{3}\\
l_{1}k_{3}s_{23} & k_{3}l_{2}s_{3} & 0
\end{array}\right].
\]}
$A_{\tau}\left(\theta,p\right)\in\mathbb{R}^{3\times4}$ is the generalized force matrix 
{\footnotesize \[
A_{\tau}\left(\theta,p\right)=\left[\begin{array}{cccc}
0 & -1 & -l_{1}s_{1}-l_{2}s_{12}-l_{3}s_{123} & l_{1}c_{1}+l_{2}c_{12}+l_{3}c_{123}\\
0 & -1 & -l_{2}s_{12}-l_{3}s_{123} & l_{2}c_{12}+l_{3}c_{123}\\
1 & -1 & -l_{3}s_{123} & l_{3}c_{123}
\end{array}\right].
\]}
The values of $p$ are bounded by the additive uncertainties in Table \ref{tab:Nominal}, so that element-wise inequalities $ p_{\min} \leq p \leq p_{\max}$ hold, and their nominal values are referred to as $\bar{p}$.
\begin{table}[b]
	\caption{Nominal Parameters of the System and their Uncertainties\label{tab:Nominal}}	
	\centering{}
	\begin{tabular}{ccccc}
		\toprule 
		Link & $m_{i}\:\left[kg\right]$ & $I_{i}\:\left[kg\cdot m^{2}\right]$ & $l_{i}\:\left[m\right]$ & $l_{ci}\:\left[m\right]$\tabularnewline
		\midrule
		\midrule 
		1 & $9.68\pm0.1$ & $1.16\pm0.1$ & $0.53\pm0.01$ & $\frac{l_{1}}{2}\pm0.01$\tabularnewline
		\midrule 
		2 & $12.59\pm0.1$ & $0.52\pm0.1$ & $0.41\pm0.01$ & $\frac{l_{2}}{2}\pm0.01$\tabularnewline
		\midrule 
		3 & $44.57\pm0.1$ & $2.56\pm0.1$ & $0.52\pm0.01$ & $\frac{l_{3}}{2}\pm0.01$\tabularnewline
		\bottomrule
	\end{tabular}
\end{table}

\section{MOTION PLANNING}

Biomechanical studies measure the kinematics of the CoM of the human body to classify and assess dynamic balance of the STS movement \cite{Fujimoto2012} rather than joint angles. Therefore, considering $\theta_{2}$, and the position coordinates of the CoM of the three-link planar robot in its inertial frame $\left(x_{CoM},y_{CoM}\right)$, we define ${\small z:=\left[\begin{array}{ccc} \theta_{2}; & x_{CoM}; & y_{CoM}\end{array}\right]}$ and plan the STS motion over the finite time horizon ${\small t\in\left[0,t_{f}\right]}$ with reference trajectories
{\small \begin{align}\begin{split}
	\bar{\theta}_{2}\left(t\right) & =  \bar{\theta}_{2}\left(0\right)+\left(\bar{\theta}_{2}\left(t_{f}\right)-\bar{\theta}_{2}\left(0\right)\right)\Phi_{1}\left(t,t_{f}\right),\\
	\bar{x}_{CoM}\left(t\right)    & =  \bar{x}_{CoM}\left(0\right)+\left(\bar{x}_{CoM}\left(t_{f}\right)-\bar{x}_{CoM}\left(0\right)\right)\Phi_{2}\left(t,t_{f}\right),\\
	\bar{y}_{CoM}\left(t\right)    & =  \bar{y}_{CoM}\left(0\right)+\left(\bar{y}_{CoM}\left(t_{f}\right)-\bar{y}_{CoM}\left(0\right)\right)\Phi_{3}\left(t,t_{f}\right),
	\end{split}\label{eq:MotionPlanning}\end{align}}where ${\small \Phi_{i}\left(t,t_{f}\right)}$ with $i=1,2,3$ are polynomial functions satisfying ${\small \Phi_{i}\left(0,t_{f}\right)=0}$ and ${\small \Phi_{i}\left(t_{f},t_{f}\right)=1}$. This rest-to-rest maneuver formulation is taken from \cite{Sira-Ramirez2004}.

Relying on kinematic equations, we showed in \cite{Narvaez-Aroche2017} that for feasible and realistic STS movements excluding the vertical position ($\theta_{1}=\nicefrac{\pi}{2}$, $\theta_{2}=\theta_{3}=0$), a transformation of the form 
{\small \begin{equation}
\left[
\bar{\theta}\left(t\right); \;\; \dot{\bar{\theta}}\left(t\right); \;\; \ddot{\bar{\theta}}\left(t\right)\right]=h\left(\bar{z}\left(t\right),\dot{\bar{z}}\left(t\right),\ddot{\bar{z}}\left(t\right),\bar{p}\right)
\label{eq:z2theta}\end{equation}}exists; so that once $\dot{\bar{z}}$ and $\ddot{\bar{z}}$ are computed from (\ref{eq:MotionPlanning}), the reference trajectories in the $z$ space can be mapped into $\theta$.

We take a computed torque \cite{Slotine1991} approach for obtaining the reference trajectories $\bar{u}\left(t\right)$. Since the system of equations in (\ref{eq:EulerLagrange}) is underdetermined,  we solve, at every $t\in\left[0,t_{f}\right]$, a control allocation problem \cite{Johansen2013} with the constrained least-squares program {\small \begin{align}
	\bar{u}\left(t\right)= & \:\underset{\xi\in\mathbb{R}^4}{\arg\min} \quad \frac{1}{2}\left\Vert W_{u}\:\xi\right\Vert _{2}^{2}\label{eq:Allocation}\\
	                       & \mathrm{subject\: to\:}   A_{\tau}\left(\bar{\theta}\left(t\right),\bar{p}\right)\xi=M\left(\bar{\theta}\left(t\right),\bar{p}\right)\ddot{\bar{\theta}}\left(t\right)+F\left(\bar{\theta}\left(t\right),\dot{\bar{\theta}}\left(t\right),\bar{p}\right)\nonumber\\
	                       & \quad \quad \quad \quad u_{\min}\leq\xi\leq u_{\max},\nonumber
\end{align}}where $W_{u}\in\mathbb{R}^{4\times4}$ and $u_{\min},u_{\max}\in\mathbb{R}^{4}$ are user-specified weights and box constraints, respectively.

\section{FINITE TIME HORIZON LQR CONTROLLER}

The Euler-Lagrange equations must be linearized in order to design an LQR controller. Define $x\in\mathbb{R}^6$ as {\small $ x:=\left[\,\theta; \:\dot{\theta}\right] $}, from (\ref{eq:EulerLagrange}), the dynamics of the three-link planar robot are 
{\small \begin{align*}
	\dot{x}\left(t\right) & = \left[\begin{array}{c}
		\dot{\theta}\left(t\right)\\
		M^{-1}\left(\theta\left(t\right),p\right)\left(A_{\tau}\left(\theta\left(t\right),p\right)u\left(t\right)-F\left(\theta\left(t\right),\dot{\theta}\left(t\right),p\right)\right)
	\end{array}\right]\\
	& =: f\left(x\left(t\right),p,u\left(t\right)\right)
\end{align*}}Thus, the state deviation variables $\delta_x\left(t\right)=x\left(t\right)-\bar{x}\left(t\right)$ satisfy {\small \[
\dot{\delta}_{x}\left(t\right):=f\left(x\left(t\right),p,u\left(t\right)\right)-f\left(\bar{x}\left(t\right),\bar{p},\bar{u}\left(t\right)\right),
\]}which can be approximated with a first order Taylor series expansion of $f\left(x\left(t\right),p,u\left(t\right)\right)$ about $\bar{x}\left(t\right)$, $\bar{p}$ and $\bar{u}\left(t\right)$: {\small \begin{align}
	\dot{\delta}_{x}\left(t\right) & \approx \left.\frac{\partial f\left(x,p,u\right)}{\partial x}\right|_{\scriptsize {\begin{array}{l}
			x=\bar{x}\left(t\right)\\
			p=\bar{p}\\
			u=\bar{u}\left(t\right)
	\end{array}}}\left(x\left(t\right)-\bar{x}\left(t\right)\right)
	+\left.\frac{\partial f\left(x,p,u\right)}{\partial p}\right|_{\scriptsize {\begin{array}{l}
			x=\bar{x}\left(t\right)\\
			p=\bar{p}\\
			u=\bar{u}\left(t\right)
	\end{array}}}\left(p-\bar{p}\right)
	+\left.\frac{\partial f\left(x,p,u\right)}{\partial u}\right|_{\scriptsize {\begin{array}{l}
			x=\bar{x}\left(t\right)\\
			p=\bar{p}\\
			u=\bar{u}\left(t\right)
	\end{array}}}\left(u\left(t\right)-\bar{u}\left(t\right)\right)\nonumber\\
	& = A\left(t\right)\delta_{x}\left(t\right)+B_{1}\left(t\right)\delta_{p}+B_{2}\left(t\right)\delta_{u}\left(t\right).\label{eq:LTV}
\end{align}}
From \cite{Athans1966}, for unconstrained $\delta_{u}\left(t\right)$, symmetric matrices $Q,S\succeq0$ and $R\succ0$, the optimal control of the stabilizable LTV system in (\ref{eq:LTV}) with $\delta_{x}\left(t\right)$ as output, and quadratic cost 
%{\small %\begin{align}
%	\begin{split}
%	\dot{\delta}_{x}\left(t\right) & = 		A\left(t\right)\delta_{x}\left(t\right)+B_{1}\left(t\right)\delta_{p}+B_{2}\left(t\right)\delta_{u}\left(t\right)\nonumber \\
%	y\left(t\right) & = \delta_{x}\left(t\right)\label{eq:LQRLTV}
%	\end{split}
%\end{align}}
{\footnotesize \[
J_{LQR}=\frac{1}{2}\delta_{x}^{\top}\left(t_{f}\right)S\delta_{x}\left(t_{f}\right)+\frac{1}{2}\int_{0}^{t_{f}}\left( \delta_{x}^{\top}\left(t\right)Q\delta_{x}\left(t\right)+\delta_{u}^{\top}\left(t\right)R\delta_{u}\left(t\right)\right) \, dt
\]}exists, is unique, time varying, and is given by {\small \begin{align} \begin{split}
	\delta_{u}\left(t\right) & = -R^{-1}B_{2}^{\top}\left(t\right)P\left(t\right)\delta_{x}\left(t\right)\\
	& = -K_{LQR}\left(t\right)\delta_{x}\left(t\right), \label{eq:LQRgain}
\end{split} \end{align}}where, considering the boundary condition $P\left(t_{f}\right)=S$, $P\left(t\right)$ is the solution of the Riccati matrix differential equation 
{\small \begin{equation}
\dot{P}\left(t\right)=-P\left(t\right)A\left(t\right)-A^{\top}\left(t\right)P\left(t\right)+P\left(t\right)B_{2}\left(t\right)R^{-1}B_{2}^{\top}\left(t\right)P\left(t\right)-Q.\label{eq:Ric}
\end{equation}}
The closed-loop nonlinear dynamics of the powered lower limb orthosis performing the STS movement under state feedback control become
\begin{equation}
	\dot{x}\left(t\right)=f\left(x\left(t\right),p,\bar{u}\left(t\right)-K_{LQR}\left(t\right)\left(x\left(t\right)-\bar{x}\left(t\right)\right)\right). \label{eq:Nonlinear}
\end{equation}
We use the numerical tools documented in \cite{Moore2015} to solve for $K_{LQR}\left(t\right)$ over $t\in\left[0,t_f\right]$.

\section{FINITE TIME HORIZON \\ ROBUST PERFORMANCE ANALYSIS}

The finite time horizon 2-norm of a signal $ v:\left[0,T\right] \rightarrow \mathbb{R}^{n}$ is defined as $ \left\Vert v\right\Vert _{2,\left[0,T\right]}:={\textstyle \left(\int_{0}^{T}v\left(t\right)^{\top}v\left(t\right)\,dt\right)^{\nicefrac{1}{2}}} $. If $ \left\Vert v\right\Vert _{2,\left[0,T\right]} $ is finite then $ v \in \mathcal{L}_{2}\left[0,T\right]$ \cite{Seiler2017}.  

\begin{figure}[H]
	\begin{centering}
	\includegraphics[width=9.5cm]{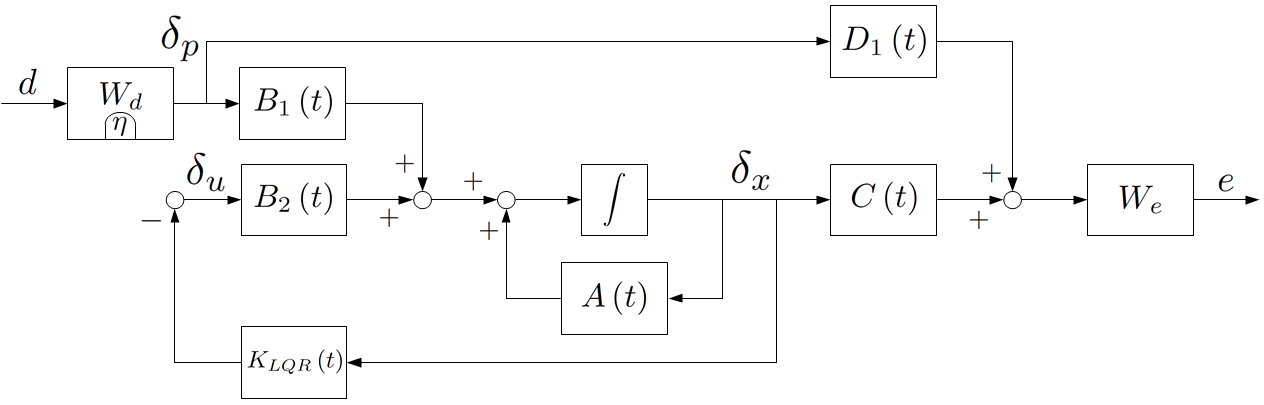}
	\par\end{centering}
	\caption{Extended LTV system for robust performance analysis.\label{fig:LTV} }
\end{figure}

In order to define a metric that captures the effect of constant, uncertain parameters on the performance of \[
\dot{\delta}_{x}\left(t\right)= \left(A\left(t\right)-B_{2}\left(t\right)K_{LQR}\left(t\right)\right)\delta_{x}\left(t\right)+B_{1}\left(t\right)\delta_{p},\label{eq:LQRLTV}
\]and to minimize the  deviation of the variables of interest from their reference trajectories $z\left(t\right)-\bar{z}\left(t\right)$ and $\dot{z}\left(t\right)-\dot{\bar{z}}\left(t\right)$, we construct the extended LTV system in Fig. \ref{fig:LTV}, where the input signal $d\in\mathbb{R}^{12}$ is assigned to be drawn from $ \mathcal{L}_{2}\left[0,T\right] $ and the output $e\in\mathbb{R}^6$ is the deviation of the variables of interest weighted by $W_{e}\in\mathbb{R}^{6\times6}$. 
In terms of $x$ and $p$, the variables of interest are computed from the kinematic equations of the three-link planar robot
{\small \begin{align*}
	\left[\begin{array}{c}
		z \\ \dot{z}
	\end{array}\right] = \left[\begin{array}{c}
	\theta_2 \\ x_{CoM} \\ y_{CoM} \\ \dot{\theta}_2 \\ \dot{x}_{CoM} \\ \dot{y}_{CoM} 
	\end{array}\right] & = \left[\begin{array}{c}
		\theta_{2}\\
		k_{0}\left(k_{1}c_{1}+k_{2}c_{12}+k_{3}c_{123}\right)\\
		k_{0}\left(k_{1}s_{1}+k_{2}s_{12}+k_{3}s_{123}\right)\\
		\dot{\theta}_{2}\\
		-\dot{\theta}_{1}y_{CoM}-\dot{\theta}_{2}k_{0}\left(k_{2}s_{12}+k_{3}s_{123}\right)-\dot{\theta}_{3}k_{0}k_{3}s_{123}\\
		\dot{\theta}_{1}x_{CoM}+\dot{\theta}_{2}k_{0}\left(k_{2}c_{12}+k_{3}c_{123}\right)+\dot{\theta}_{3}k_{0}k_{3}c_{123}
	\end{array}\right]\\
	& =: \zeta\left(x,p\right),
\end{align*}}The deviation from their desired trajectories \[
\delta_{\zeta}\left(t\right) := \zeta\left(x\left(t\right),p\right)-\zeta\left(\bar{x}\left(t\right),\bar{p}\right)
\] is approximated by a first order Taylor series expansion of $\zeta\left(x,p\right)$ about $\bar{x}\left(t\right)$ and $\bar{p}$: \begin{equation}
\delta_{\zeta}\left(t\right)\approx C\left(t\right)\delta_{x}\left(t\right)+D_{1}\left(t\right)\delta_{p},\label{eq:zetadevlin}
\end{equation}with 
{\small \begin{eqnarray*}
	C(t):= \left.\frac{\partial\zeta\left(x,p\right)}{\partial x}\right|_{\scriptsize {\begin{array}{l}
			x=\bar{x}\left(t\right)\\
			p=\bar{p}
	\end{array}}}& & 
	D_{1}\left(t\right):= \left.\frac{\partial\zeta\left(x,p\right)}{\partial p}\right|_{\scriptsize {\begin{array}{l}
			x=\bar{x}\left(t\right)\\
			p=\bar{p}.
	\end{array}}}
\end{eqnarray*}}

Although $ \delta_p $ is an unknown constant, in this section it is treated as a time-varying signal. Specifically, we choose a strictly proper system $W_{D}$: {\small \begin{align}
	\begin{split}
	\dot{\eta}\left(t\right) & =  A_{d}\eta\left(t\right)+B_{d}d\left(t\right) \\
	\delta_{p}\left(t\right) & =  C_{d}\eta\left(t\right)\label{eq:Wd1},
	\end{split}
\end{align}}with $A_d,B_d,C_d\in\mathbb{R}^{12\times12}$, whose output models $ \delta_{p}\left(t\right) $.

Defining {\small \begin{align*}
\bar{A}\left(t\right)&:=\left[\begin{array}{cc}
A\left(t\right)-B_{2}\left(t\right)K_{LQR}\left(t\right) & B_{1}\left(t\right)C_{d}\\
0 & A_{d} \end{array}\right]\\
\bar{B}\left(t\right)&:=\left[\begin{array}{c}
B_{1}\left(t\right)D_{d}\\
B_{d}\end{array}\right] \quad \bar{C}\left(t\right):=\left[\begin{array}{cc}
W_{e}C\left(t\right) & W_{e}D_{1}\left(t\right)C_{d}\end{array}\right],
\end{align*}}the state space realization of the extended LTV system in Fig. \ref{fig:LTV} is{\small \begin{align}\begin{split}
\left[\begin{array}{c}
\dot{\delta}_{x}\left(t\right)\\
\dot{\eta}\left(t\right)
\end{array}\right] & = \bar{A}\left(t\right)\left[\begin{array}{c}
\delta_{x}\left(t\right)\\
\eta\left(t\right)
\end{array}\right]+\bar{B}\left(t\right)d\left(t\right)\\
e\left(t\right) & = \bar{C}\left(t\right)\left[\begin{array}{c}
\delta_{x}\left(t\right)\\
\eta\left(t\right)
\end{array}\right],\label{eq:WeightedLTV}
	\end{split}\end{align}}and its \textit{finite-horizon} $\mathcal{L}_{2}$\textit{-to-Euclidean gain} \cite{Seiler2017} is \[
\gamma_{\left[0,T\right]}:=\sup\left\{ \left.\frac{\left\Vert e\left(T\right)\right\Vert _{2}}{\left\Vert d\right\Vert _{2,\left[0,T\right]}}\right|\left[\begin{array}{c}
\delta_{x}\left(0\right)\\\eta\left(0\right)\end{array}\right]=0,0\neq d\in\mathcal{L}_{2}\left[0,T\right]\right\}. 
\]
The most important feature of the LQR controller is to guarantee a safe transition to the standing position in the presence of parameter uncertainty by the end of the finite time horizon. For this purpose, we choose as robust performance metric:   {\small \begin{equation}
J_{RP}=\left(1-\alpha\right)\gamma_{\left[0,t_{m}\right]}+\alpha \gamma_{\left[0,t_{f}\right]}\label{eq:JRP},
\end{equation}}where $\alpha\in\left[0,1\right]$ weights the induced gains at an intermediate time $t_{m}$ and a final time $ t_f $. 

To calculate the induced gains we use the computational approach from \cite{Seiler2017} and the toolbox documented in \cite{Moore2015}.  

\section{WEIGHT SELECTION VIA ROBUST PERFORMANCE METRIC}

Since $J_{RP}$ depends on the choice of $Q$, $R$, $S$ in the design of the LQR controller (\ref{eq:LQRgain})-(\ref{eq:Ric}), we choose finite sets of candidates $\mathcal{Q}\subset\mathbb{R}^{6\times6}$, $\mathcal{R}\subset\mathbb{R}^{4\times4}$ and $\mathcal{S}\subset\mathbb{R}^{6\times6}$, draw an element from each of them, obtain their corresponding LQR controller and compute $J_{RP}$. Taking a brute force approach, the weight matrices
\begin{equation}
	Q^{\star},R^{\star},S^{\star}=\underset{Q\in\mathcal{Q},R\in\mathcal{R},S\in\mathcal{S}}{\arg\min}J_{RP}\left(Q,R,S\right)\label{eq:WeightSelection}
\end{equation}characterize the best gain $ K_{LQR}^{\star}\left(t\right) $ from the pool of candidates, relative to our robust performance metric.

\section{SIMULATION OF TWO STS MOVEMENTS}

The movement STS 1 starts with the shank and torso segments parallel to the vertical, and the thigh segment parallel to the horizontal by setting $\theta_{1}\left(0\right)=90\text{\textdegree}$, $\theta_{2}\left(0\right)=-90\text{\textdegree}$ and $\theta_{3}\left(0\right)=90\text{\textdegree}$ ($\bar{x}_{CoM}\left(0\right)=0.309$ and $\bar{y}_{CoM}\left(0\right)=0.6678$). STS 2 vertically aligns the CoM and the ankle joint prior to seat-off with the initial conditions $\theta_{1}\left(0\right)=120\text{\textdegree}$, $\theta_{2}\left(0\right)=-120\text{\textdegree}$, $\theta_{3}\left(0\right)=110.87\text{\textdegree}$ ($\bar{x}_{CoM}\left(0\right)=0$ and $\bar{y}_{CoM}\left(0\right)=0.590\left[m\right]$). For both movements, the final standing configuration places the CoM directly above the origin of the inertial frame with the values $\bar{\theta}_{2}\left(t_{f}\right)=-5\text{\textdegree}$, $\bar{x}_{CoM}\left(t_{f}\right)=0$ and $\bar{y}_{CoM}\left(t_{f}\right)=0.974\left[m\right]$. STS 1 and STS 2 are respectively referred in the biomechanical literature as dynamic and quasi-static strategies \cite{Galli2008}. To complete the design of the rest-to-rest maneuvers in terms of $ \bar{z}\left(t\right) $, $ \dot{\bar{z}}\left(t\right) $ and $ \ddot{\bar{z}}\left(t\right) $ with (\ref{eq:MotionPlanning}), define $\Phi_{i}\left(t,t_{f}\right):=-2\frac{t^{3}}{t_{f}^{3}}+3\frac{t^{2}}{t_{f}^{2}}$, which is the only cubic polynomial satisfying $\dot{\Phi}\left(0,t_{f}\right)=\dot{\Phi}\left(t_{f},t_{f}\right)=0$ (and $\Phi\left(0,t_{f}\right)=0$ and $\Phi\left(t_{f},t_{f}\right)=1$). Due to the lack of data on comfortable STS duration for subjects with complete spinal cord injuries, we picked one reported in \cite{Janssen2008} for stroke patients; leading to a simulation time of $ t_{f}=3.5\left[s\right] $ for both movements. 

The reference state trajectories for each movement are determined from (\ref{eq:z2theta}). When solving for $ \bar{u}\left(t\right) $ in (\ref{eq:Allocation}), it is enforced that the contributions from $\tau_{1}\left(t\right)$, $\tau_{2}\left(t\right)$ and $F_{y}\left(t\right)$ outweigh $F_{x}\left(t\right)$ by considering $W_u=\operatorname{diag}\left(\left[ 1 \quad 1 \quad 10 \quad 1\right]\right)$ and, because the user always pushes the crutches down to propel upwards, the constraint $F_{y}\left(t\right)\geq0$ is imposed. All other inputs are unconstrained.

After numerically computing the linearizations in (\ref{eq:LTV}) and (\ref{eq:zetadevlin}), the formulation of the extended LTVs in (\ref{eq:WeightedLTV}) is completed specifying $ W_d $ with:{\small \begin{align*}
A_{d} & = -a\mathcal{I}_{12} \hspace{2.5cm} B_{d} = \mathcal{I}_{12}\\
C_{d} & = a\operatorname{diag}\left(\frac{p_{max}-p_{min}}{2}\right)  %& D_{d} & = \mathbf{0}
\end{align*}}For a bandwidth of $50\left[Hz\right]$ we set $a=100\pi$ and the deviations of the variables of interest are penalized with $ W_{e}=\operatorname{diag}\left(\left[1 \quad 1 \quad 1 \quad 10 \quad 10 \quad 10\right]\right) $. Finally, $ \alpha:= 0.7 $ and $t_{m}:= 2\left[s\right]$.

The LQR weight matrices candidates in (\ref{eq:WeightSelection}) were limited to be diagonal, positive definite matrices, with entries sampled from a Latin Hypercube of 1350 experiments. The entries of $ \mathcal{Q} $ and $ \mathcal{S} $ are in $ \left(0,10^4\right) $, while the entries of $ \mathcal{R} $ belong to $ \left(0,1\right) $. The $ \arg\min $ triplets in (\ref{eq:WeightSelection}) for tracking each of the STS movements are: 
%{\small \begin{align*}
%	Q^{\star}_{1} & = \operatorname{diag}\left(\left[\begin{array}{ccc}
%	3237 & 5534 & 6546\ldots\end{array}\right.\right.\\
%	& \left.\left.\begin{array}{cccccc}
%	& & & \;\; 7918 & 4003 & 8516\end{array}\right]\right)\\
%	R^{\star}_{1} & = \operatorname{diag}\left(\left[\begin{array}{cccc}
%	0.3659 & 0.0155 & 0.1433 & 0.1553\end{array}\right]\right)\\
%	S^{\star}_{1} & = \operatorname{diag}\left(\left[\begin{array}{ccc}
%	1068 & 5396 & 1324\ldots\end{array}\right.\right.\\
%	& \left.\left.\begin{array}{ccccccc}
%	& & & \;\;9467 & 3975 & 5819\end{array}\right]\right)\\
%	Q^{\star}_{2} & = \operatorname{diag}\left(\left[\begin{array}{ccc}
%	3766 & 9550 & 2932\ldots\end{array}\right.\right.\\
%	&   \left.\left.\begin{array}{cccccc}
%	& & & \;\;	8378 & 9552 & 9242\end{array}\right]\right)\\
%	R^{\star}_{2} & = \operatorname{diag}\left(\left[\begin{array}{cccc}
%	0.1119 & 0.0252 & 0.3600 & 0.3045\end{array}\right]\right)\\
%	S^{\star}_{2} & = \operatorname{diag}\left(\left[\begin{array}{ccc}
%	9565 & 820 & 5316\ldots\end{array}\right.\right.\\
%	&   \left.\left.\begin{array}{cccccc}
%	& & & \;\; 5779 & 6083 & 8877\end{array}\right]\right)
%	\end{align*}}
{\small \begin{align*}
	Q^{\star}_{1} & = \operatorname{diag}\left(\left[3237 \quad 5534 \quad 6546 \quad 7918 \quad 4003 \quad 8516\right]\right)\\
	R^{\star}_{1} & = \operatorname{diag}\left(\left[0.3659 \quad 0.0155 \quad 0.1433 \quad 0.1553\right]\right)\\
	S^{\star}_{1} & = \operatorname{diag}\left(\left[1068 \quad 5396 \quad 1324 \quad 9467 \quad 3975 \quad 5819\right]\right)\\
	Q^{\star}_{2} & = \operatorname{diag}\left(\left[3766 \quad 9550 \quad 2932 \quad 8378 \quad 9552 \quad 9242\right]\right)\\
	R^{\star}_{2} & = \operatorname{diag}\left(\left[0.1119 \quad 0.0252 \quad 0.3600 \quad 0.3045\right]\right)\\
	S^{\star}_{2} & = \operatorname{diag}\left(\left[9565 \quad 820 \quad 5316 \quad 5779 \quad 6083 \quad 8877\right]\right)
\end{align*}}Their corresponding performance metrics are $ J_{RP1}=0.1571 $ and $ J_{RP2}=0.1553 $.

\begin{figure}[H]
	\centering
	\subfloat[Angular position of link 1 relative to the horizontal.\label{fig:theta1} ]{\includegraphics[width=7.5cm]{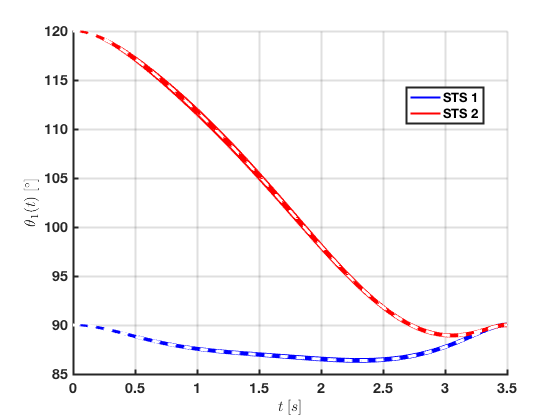}} \hspace{1.5cm}
	\subfloat[Angular position of link 2 relative to link 1.\label{fig:theta2} ]{\includegraphics[width=7.5cm]{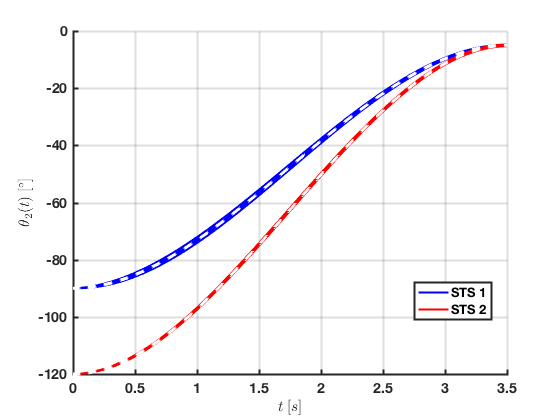}}
	\\
	\subfloat[Angular position of link 3 relative to link 2.\label{fig:theta3}]{ \includegraphics[width=7.5cm]{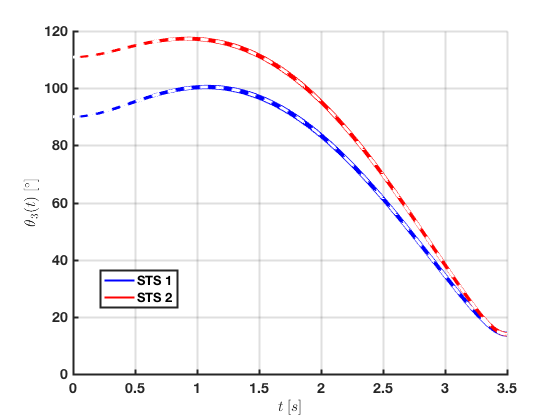}} \hspace{1.5cm}
	\subfloat[Angular velocity of link 1.\label{fig:omega1}]{ \includegraphics[width=7.5cm]{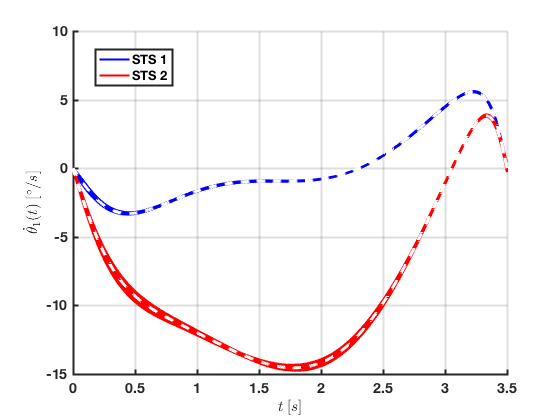}}
	\\
	\subfloat[Angular velocity of link 2.\label{fig:omega2}]{ \includegraphics[width=7.5cm]{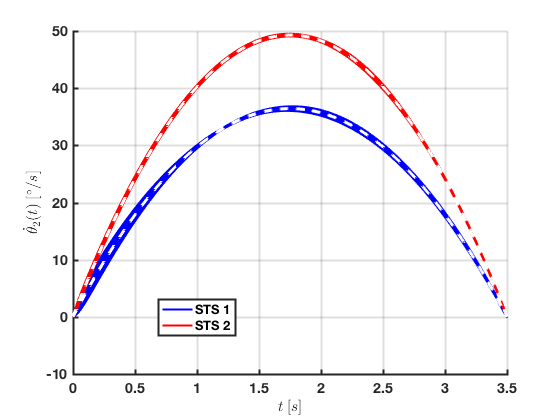}} \hspace{1.5cm}
	\subfloat[Angular velocity of link 3.\label{fig:omega3}]{ \includegraphics[width=7.5cm]{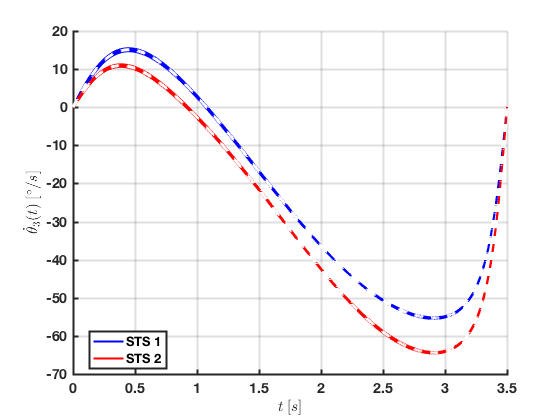}}
	\caption{State of the nonlinear system under finite time robust control for two relevant STS movements.\label{fig:State}}
\end{figure}

\begin{figure}[H]
\centering
\subfloat[Torque applied at the hips by the powered lower limb orthosis.\label{fig:tau1}]{ \includegraphics[width=7.5cm]{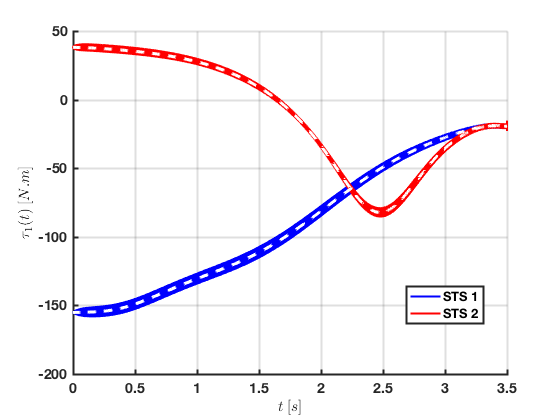}} \hspace{1.5cm}
\subfloat[Torque at the shoulders of the user.\label{fig:tau2}]{ \includegraphics[width=7.5cm]{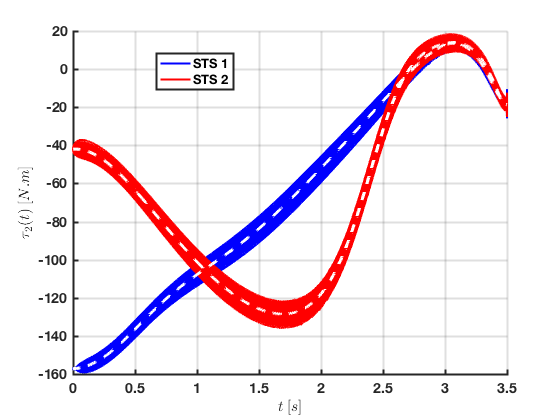}}
\\
\subfloat[Horizontal force at the shoulders of the user.\label{fig:Fx}]{ \includegraphics[width=7.5cm]{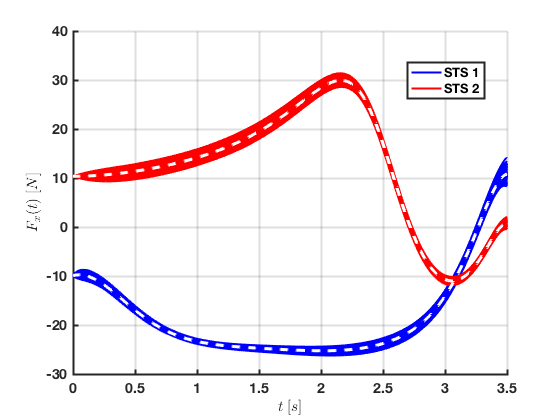}} \hspace{1.5cm}
\subfloat[Vertical force at the shoulders of the user.\label{fig:Fy}]{ \includegraphics[width=7.5cm]{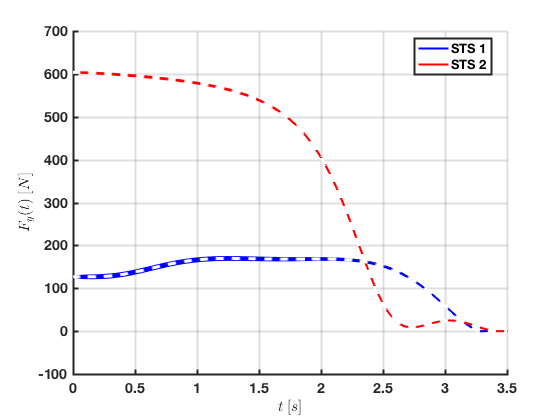}}
\caption{Inputs from the finite time horizon LQR controllers for two relevant STS movements.\label{fig:Input}}
\end{figure}

\begin{figure}[H]
\centering
\subfloat[Position trajectories of the three-link robot CoM.\label{fig:XY}]{ \includegraphics[width=7.5cm]{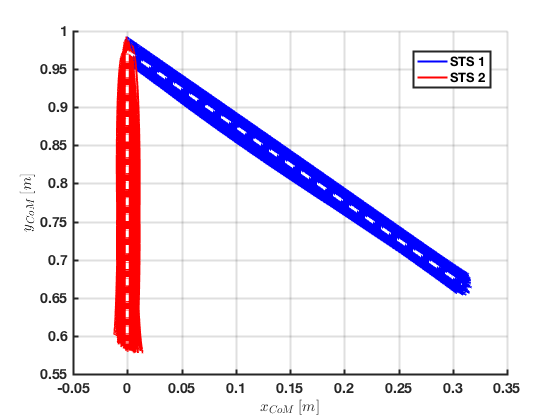}} \hspace{1.5cm}
\subfloat[Velocity trajectories of the three-link robot CoM.\label{fig:VXVY}]{ \includegraphics[width=7.5cm]{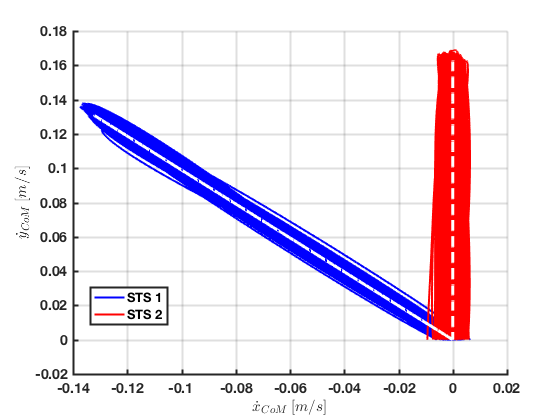}}
\caption{Center of Mass trajectories achieved with the finite time robust control of two relevant STS movements.\label{fig:CoM}}
\end{figure}

The simulations of the full nonlinear system under finite time robust control (\ref{eq:Nonlinear}) with $ K_{LQR1}^{\star}\left(t\right) $ for STS 1 and $ K_{LQR2}^{\star}\left(t\right) $ for STS 2 are in Figs. \ref{fig:State}-\ref{fig:CoM}. The dashed lines represent the evolution of the variables when $ p=\bar{p} $. The collection of continuous lines represents the evolution of the variables for the same 200 sets of parameter values randomly chosen in \cite{Narvaez-Aroche2017}. 
Figures \ref{fig:theta1}-\ref{fig:Fy} verify that tracking of the reference state and input trajectories dramatically improved with respect to the results of the scheme based on feedback linearization and control allocation in \cite{Narvaez-Aroche2017}, regardless of the STS maneuver, essentially overlapping with the nominal trajectories in some of the plots. Even though it might seem unrealistic that users would be able to exactly mimic the inputs from the LQR controllers in Fig. \ref{fig:Input}, we believe that it should be easier for them, with training, to achieve a good interaction with the orthoses to consistently execute safe STS maneuvers if their required sequences of actions do not vary much despite of parameter changes, e.g. due to daily weight fluctuation, mechanical wear of the braces or links. 
Note that the variation of the lengths and masses of the links from their nominal values causes the initial conditions for $ x_{CoM} $ and $ y_{CoM} $ in  Fig. \ref{fig:XY} not being the same across simulations, despite of each STS movement having a single initial condition for $ x $. In the same figure, it is interesting to observe how the selected LQR controller for STS 2, whose performance metric is slightly smaller than the one for the controller of STS 1, decreases the Euclidean norm of the position of the CoM at the end of the finite horizon, in accordance to the induced gain at $ t_f $ in our performance metric. Given that all the final positions of the CoM are approximately aligned with the ankle (with an error less than or equal to $ 5 \left[mm\right] $ in $ x_{CoM} $) and the magnitude of its final velocity in Fig. \ref{fig:VXVY} is less than or equal to $ 1 \left[\nicefrac{cm}{s}\right] $, there is no risk of sit-back or step failures during the STS movements with the proposed control technique.

In this initial study the parameter uncertainties were small, they will be increased in further work.

\section{CONCLUSIONS}

This paper presented a technique for performing the robust control of Sit-to-Stand movements of Powered Lower Limb Orthoses in the presence of uncertainties in their parameters. Its main tasks are:
 \begin{itemize}
 	\item Calculate reference trajectories of states and inputs from the nonlinear equations of the system, considering the nominal values of its parameters.
 	\item Obtain linear expressions for the deviations of states and other variables of interests, about reference trajectories and nominal parameters.  
 	\item Considering a finite horizon LQR controller for the Jacobian linearization of the dynamics, construct a LTV system with a perturbation signal to model the parameter variation as its input, and the deviation from the variables of interest as output.
 	\item Define a performance metric to assess the robustness of the LTV system on a finite time horizon involving the induced gains in \cite{Moore2015}. 
 	\item Choose finite sets of weight matrices candidates and search for those that lead to the LQR gain that minimizes the robust performance metric.
 	\item Use the best TV LQR gain for tracking the reference trajectories under the nonlinear dynamics of the system subject to parameter uncertainty.
 	\item If the simulations are not satisfactory, choose different sets of weight matrices candidates to keep decreasing the robust performance metric.
 \end{itemize}
This procedure can be directly applied without modification to other robotic systems that undergo finite-time trajectories. We are specifically interested in models of powered lower limb orthoses that incorporate hip \underline{and} knee actuation, accounting for more complicated architectures found in medical devices of this kind available in the market \cite{Cyberdyne2017,EksoBionics2017,ParkerHannifinCorp2017,ReWalkRobotics2017,RokiRobotics2017}.
%\pagebreak
% Additive inputs to model the parameter variation are very approximate. 
%\addtolength{\textheight}{-12cm}  % This command serves to balance the column lengths
                                  % on the last page of the document manually. It shortens
                                  % the textheight of the last page by a suitable amount.
                                  % This command does not take effect until the next page
                                  % so it should come on the page before the last. Make
                                  % sure that you do not shorten the textheight too much.

%%%%%%%%%%%%%%%%%%%%%%%%%%%%%%%%%%%%%%%%%%%%%%%%%%%%%%%%%%%%%%%%%%%%%%%%%%%%%%%%

\section*{ACKNOWLEDGMENTS}

The first author would like to thank the Consejo Nacional de Ciencia y Tecnolog\'{i}a (CONACYT), the Fulbright-Garc\'{i}a Robles program and the University of California Institute for Mexico and the United States (UC MEXUS) for the scholarships that have made possible his Ph.D. studies; and Karla Mendoza-Damken for the composition in Fig. \ref{fig:Robot}. The authors gratefully acknowledge support from the National Science Foundation under grant ECCS-1405413. Andrew Packard acknowledges the generous support from the FANUC Corporation.

%%%%%%%%%%%%%%%%%%%%%%%%%%%%%%%%%%%%%%%%%%%%%%%%%%%%%%%%%%%%%%%%%%%%%%%%%%%%%%%%
% \section*{REFERENCES}

\bibliographystyle{IEEEtran}
\bibliography{IEEEabrv,STSBiblio}

\end{document}